\documentstyle[12pt,epsf,rotate]{article}

\def\be{\begin{equation}}
\def\ee{\end{equation}}
\def\bi{\begin{itemize}}
\def\ei{\end{itemize}}
\def\bma{\begin{displaymath}}
\def\ema{\end{displaymath}}
\def\btr{\begin{tabular}}
\def\etr{\end{tabular}}
\def\bfig{\begin{figure}}
\def\efig{\end{figure}}
\def\bt{\begin{table}}
\def\et{\end{table}}
\def\ni{\noindent}

\begin{document}

\begin{center}
{\large \bf Connectivity and the Origin of Inertia}

\rule{0pt}{25pt}
L. J. Nickisch\\
{\em Mission Research Corporation}\\
{\em Monterey, California}\\
{\em nickisch@mrcmry.com}\\
{\em and}\\
Jules Mollere\\
{\em Henderson State University}\\
{\em Arkadelphia, Arkansas}\\
{\em mollerj@hsu.edu}
\end{center}

\begin{center}
\rule{0pt}{12pt}
{\bf Abstract}
\end{center}

\rule{0pt}{12pt} Newton's Second Law defines inertial mass as the
ratio of the applied force on an object to the responding
acceleration of the object (viz., $F=ma$).  Objects that exhibit
finite accelerations under finite forces are described as being
``massive'' and this mass has usually been considered to be an
innate property of the particles composing the object.  However
mass itself is never directly measured.  It is inertia, the
reaction of the object to impressed forces, that is measured.  We
show that the effects of inertia are equally well explained as a
consequence of the vacuum fields acting on massless particles
travelling in geodesic motion. In this approach, the vacuum fields
in the particle's history define the curvature of the particle's
spacetime. The metric describing this curvature implies a
transformation to Minkowski spacetime, which we call the
Connective transformation. Application of the Connective
transformation produces the usual effects of inertia when observed
in Minkowski spacetime, including hyperbolic motion in a static
electric field (above the vacuum) and uniform motion following an
impulse.  In the case of the electromagnetic vacuum fields, the
motion of the massless charge is a helical motion that can be
equated to the particle spin of quantum theory. This spin has the
properties expected from quantum theory, being undetermined until
``measured'' by applying a field, and then being found in either a
spin up or spin down state. Furthermore, the zitterbewegung of the
charge is at the speed of light, again in agreement with quantum
theory.  Connectivity has the potential for pair creation as the
Connective transformation can transform positive time intervals in
the particle spacetime to negative time intervals in Minkowski
spacetime.

\section{Introduction}

Renewed interest in explaining the origin of inertia has been
triggered by the foundational work of Rueda, Haisch, and Puthoff
[1,2].  Rather than accepting that inertial mass is simply an
innate property of matter as defined in Newton's Second Law, these
authors show that the effects of inertia can be explained as a
consequence of accelerated motion in the electromagnetic vacuum
field, the so-called Zero-Point Field (ZPF).  We develop a formal
approach to the explanation of inertia as a reaction to the vacuum
fields that simultaneously addresses aspects of quantum theory
that heretofore have been problematic from a Classical particle
point of view.  Specifically we address the quantum-theoretic
prediction that the zitterbewegung of particles occurs at the
speed of light, that particles exhibit spin, and that pair
creation can occur.

In his study of the coordinate operator in the Dirac equation,
Schr\"{o}dinger discovered microscopic oscillatory motion at the
speed of light, which he called {\em zitterbewegung} [3].  While
Dirac argued that such motion does not violate relativity or
quantum theory [4], from a Classical particle point of view, these
speed of light motions would seem to imply masslessness of the
particle.

Dirac theory also describes particle spin, and Schr\"{o}dinger
considered spin to be an orbital angular momentum that is a
consequence of the vacuum fields [3].  This view of spin was
explored further by Huang [5] and Barut and Zanghi [6].

We present a description of the motion of massless particles that
manifests inertia as a consequence of the vacuum fields.  This
approach simultaneously yields spin as an orbital angular momentum
driven by the electromagnetic vacuum (ZPF).  We derive an equation
of motion for massless charges as a limit of the Lorentz force
equation.  This equation can be viewed as a geodesic equation in a
spacetime associated with the particle, and the implied
transformation to Minkowski spacetime (the Connective
transformation) yields the effects of inertia as viewed in
Minkowski spacetime.  A massless charge is seen in Minkowski
spacetime to be driven by the ZPF in helical motions that we
interpret to be spin.  The theory has the potential to exhibit the
many-particle behavior of pair creation and annihilation as the
Connective transformation sometimes takes positive time intervals
in the particle spacetime to negative time intervals in Minkowski
spacetime.

We will formulate Connectivity in the context of electromagnetism,
where the particle is assumed to be a massless electric charge and
the vacuum fields are the ZPF.  The approach should be
generalizable to the other vacuum fields (e.g., the color fields
of the strong nuclear force).

In the following we use the metric signature $(-,+,+,+)$.

\section{Massless charge equation of motion}

A suitable equation for describing the motion of a massless charge
can be derived as the massless limit of the Lorentz force
equation,

\be m \alpha^\mu = \frac {q} {c} F^{\mu \nu} u_\nu \;\;\;,
\label{lorentz}\ee

\ni where $q$ is the charge of the particle, $c$ is the speed of
light, and $m$ is the particle mass (which we will soon take to be
zero). $F^{\mu \nu}$ is the electromagnetic field tensor of the
impressed fields, including the ZPF.  The acceleration $\alpha$
and the velocity $u$ are four-vectors.  In terms of the usual
three-space acceleration ${\bf a}$ and velocity ${\bf v}$ we have

\be m\left[\gamma^2 a^j + \gamma^4 {\bf v}\cdot {\bf a} \frac
{v^j} {c}\right]=\frac {q} {c} \gamma\left[\sum_{k=1}^3 F^{j k}v_k
- F^{j 0} c \right] \ee

\be m\gamma^4 \frac {{\bf v}\cdot {\bf a}} {c}=\frac {q} {c}
\gamma\left[\sum_{k=1}^3 F^{0 k}v_k - F^{0 0} c \right] \;\;\;,\ee

\ni where $\gamma=1/\sqrt{1-v^2/c^2}$.  These equations can be
combined to yield

\be m\gamma a^j =\frac {q} {c} \left[\sum_{k=1}^3 \left(F^{j k} -
F^{0 k} \frac {v^j} {c}\right)v_k - \left(F^{j 0}-F^{0 0}\frac
{v^j} {c}\right) c \right] \;\;\;.\ee

We take the massless limit to be that in which the mass goes to
zero as the speed of the particle becomes the speed of light
(hence $\gamma\rightarrow\infty$).  The product $m\gamma$ is
assumed to remain finite in this limit, that is,

\be \lim_{\begin{array}{c}
  m\rightarrow 0 \\
  v\rightarrow c
\end{array}} m \gamma = m_* \;\;\;.
\label{lim}\ee

\ni The mass parameter $m_*$ has the dimensions of mass, but it is
{\em not} mass.  Particles described by (\ref{lim}) move at the
speed of light and therefore must satisfy

\be {\bf v}=c {\bf n} \;\;\;, \label{vn}\ee

\ni where ${\bf n}$ is a unit vector in the direction of the
particle motion.  Accelerations therefore can only be due to
changes in the direction of the particle, not in its speed,

\be {\bf a}=c \frac {d{\bf n}} {dt} \;\;\;. \label{an}\ee

\ni We have, then,

\be m_* c \frac {d n^j} {dt} =\frac {q} {c} \left[\sum_{k=1}^3
\left(F^{j k} - F^{0 k} \frac {v^j} {c}\right)v_k - \left(F^{j
0}-F^{0 0}\frac {v^j} {c}\right) c \right] \;\;\;.\label{eqmo}\ee

\ni In terms of the electric and magnetic fields, this is

\be \frac {d{\bf {\bf n}}} {dt} = \frac {q} {m_* c}\left[{\bf {\bf
n}}\times{\bf B}- {\bf {\bf n}}\cdot{\bf E}{\bf {\bf n}}+{\bf E}
\right] \;\;\;. \label{eqmo2}\ee

Note in (\ref{eqmo2}) that only transverse forces accelerate the
particle, consistent with conditions (\ref{vn}) and (\ref{an}).
This is also consistent with the expectations of relativity. Since
it is moving at the speed of light, the particle should see a
universe Lorentz contracted to two transverse dimensions, so it
can only be accelerated by forces from the side.

Equation (\ref{eqmo}) describes the motion of a massless charge in
response to impressed electromagnetic fields. The charge moves at
a constant speed (the speed of light) with a changing direction
given by (\ref{eqmo}). When the impressed fields include the ZPF,
this motion may be regarded as Schr\"{o}dinger's zitterbewegung.
When a field above the vacuum is applied, the charge will be
observed to drift in a preferred direction in its Zitterbewegung
wander.

These effects are illustrated in Figures 1 and 2.  Figure 1 shows
the trajectory of a massless charge computed using equation
(\ref{eqmo}).  The electromagnetic fields influencing the motion
of the charge are a random realization of the ZPF (see Appendix A
for a description of how the random realization is generated) with
a superimposed uniform electric field in the vertical direction
(the driving field), which is switched off after half of the total
time duration of the simulation.  Note that the charge drifts
upward in response to the driving field.  Figure 2 shows the
region of Figure 1 near the origin.  Here we see that the ZPF
drives the charge in a pseudo-helical motion as in
Schr\"{o}dinger's orbital angular momentum explanation for spin.

\begin{figure}[h]
\hfill
\begin{minipage}[t]{2.in}
\epsfysize=2.in \centerline{\epsffile{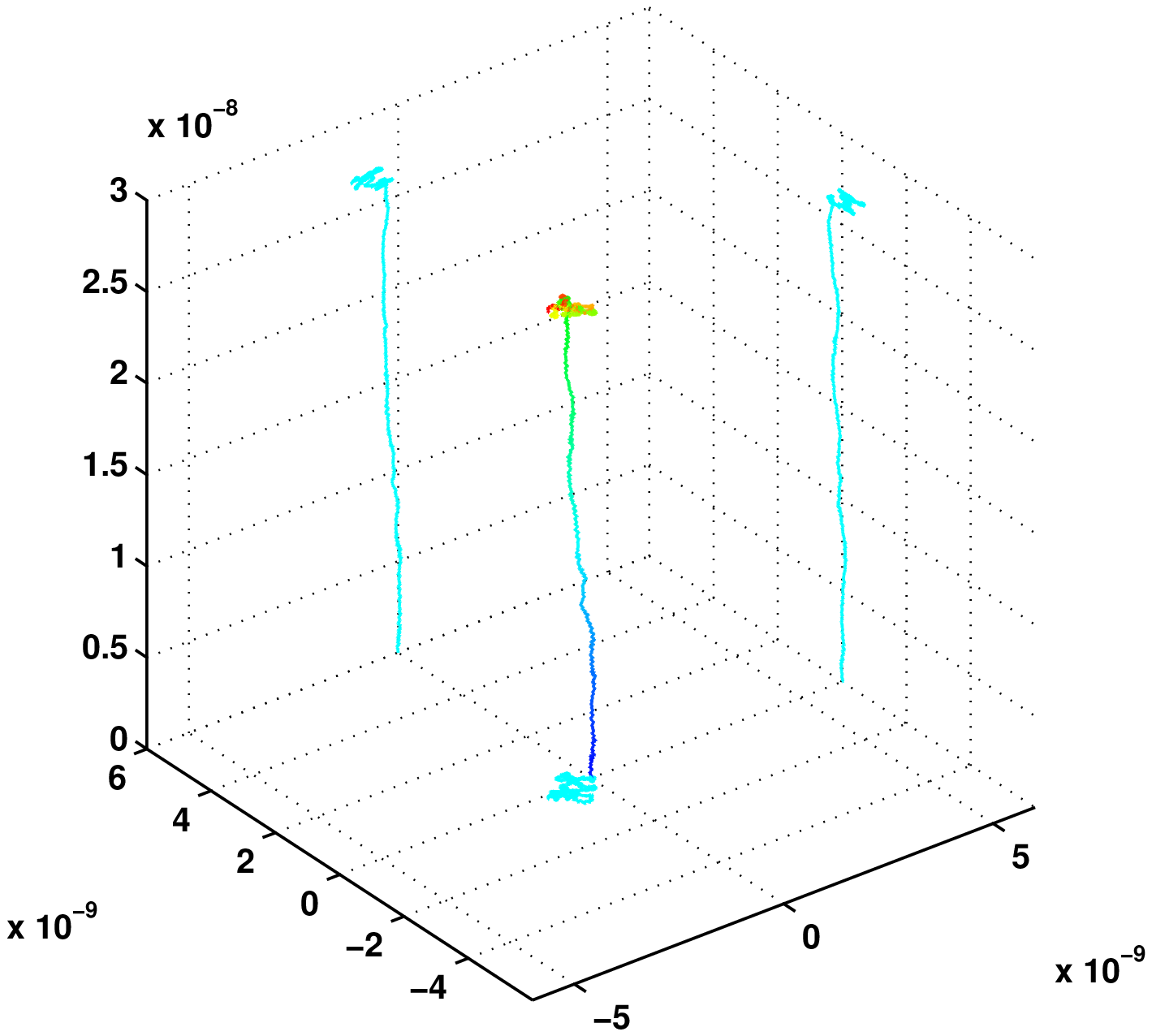} }
\begin{center}
 \begin{minipage}[t]{2.0in}
{Figure 1.  Massless charge trajectory in uniform electric field
(plus ZPF), which is then switched off after half the total time
duration.}
 \end{minipage}
\end{center}
\end{minipage}
\hfill
\begin{minipage}[t]{2.in}
\epsfysize=2.in \centerline{\epsffile{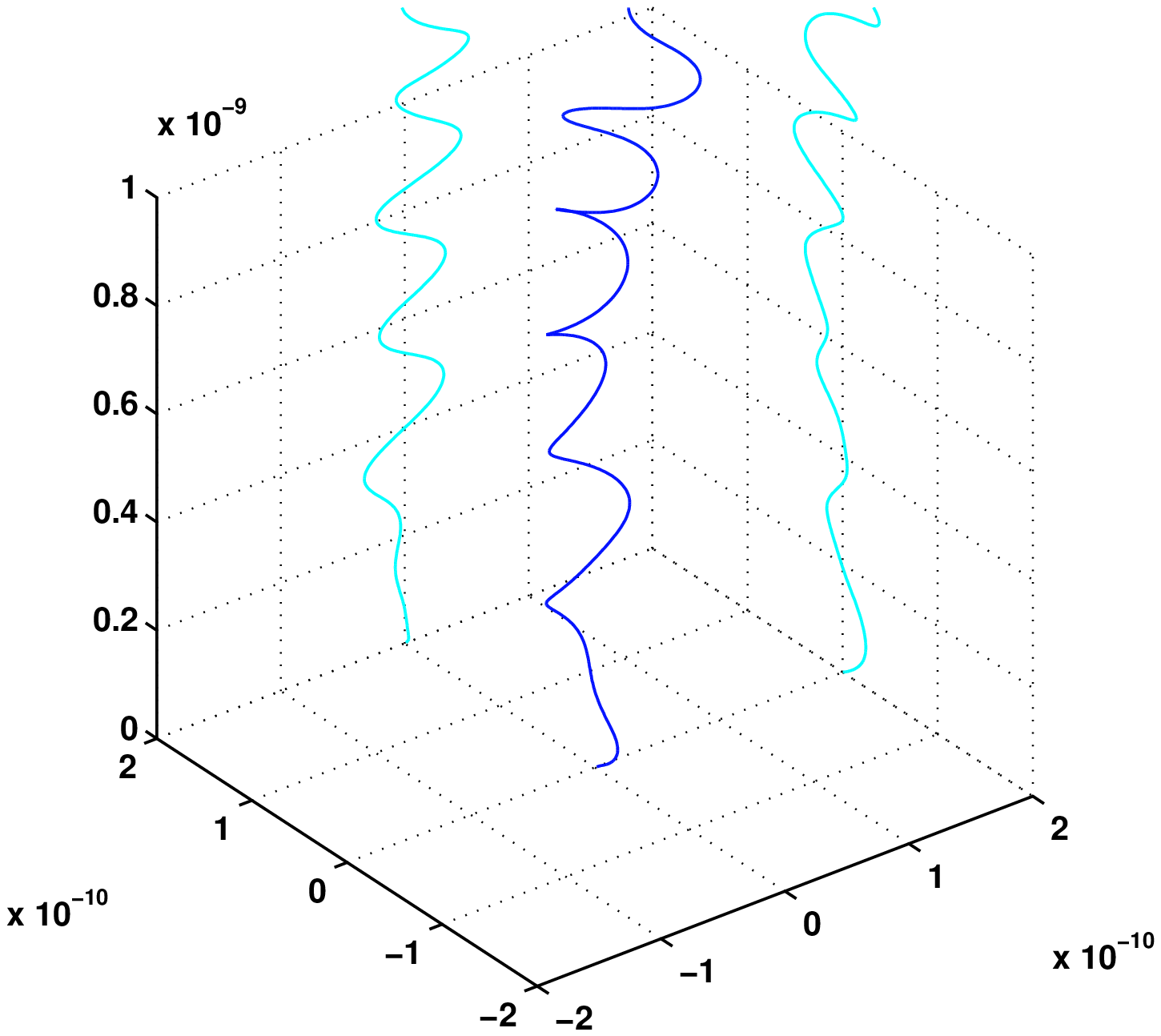} }
\begin{center}
 \begin{minipage}[t]{2.0in}
 {Figure 2. Zoomed view near the origin of Figure 1
 showing spin-like orbital motion driven by the ZPF.}
 \end{minipage}
\end{center}
\end{minipage}
\hfill
\begin{minipage}[t]{2.in}
\epsfysize=2.in \centerline{\epsffile{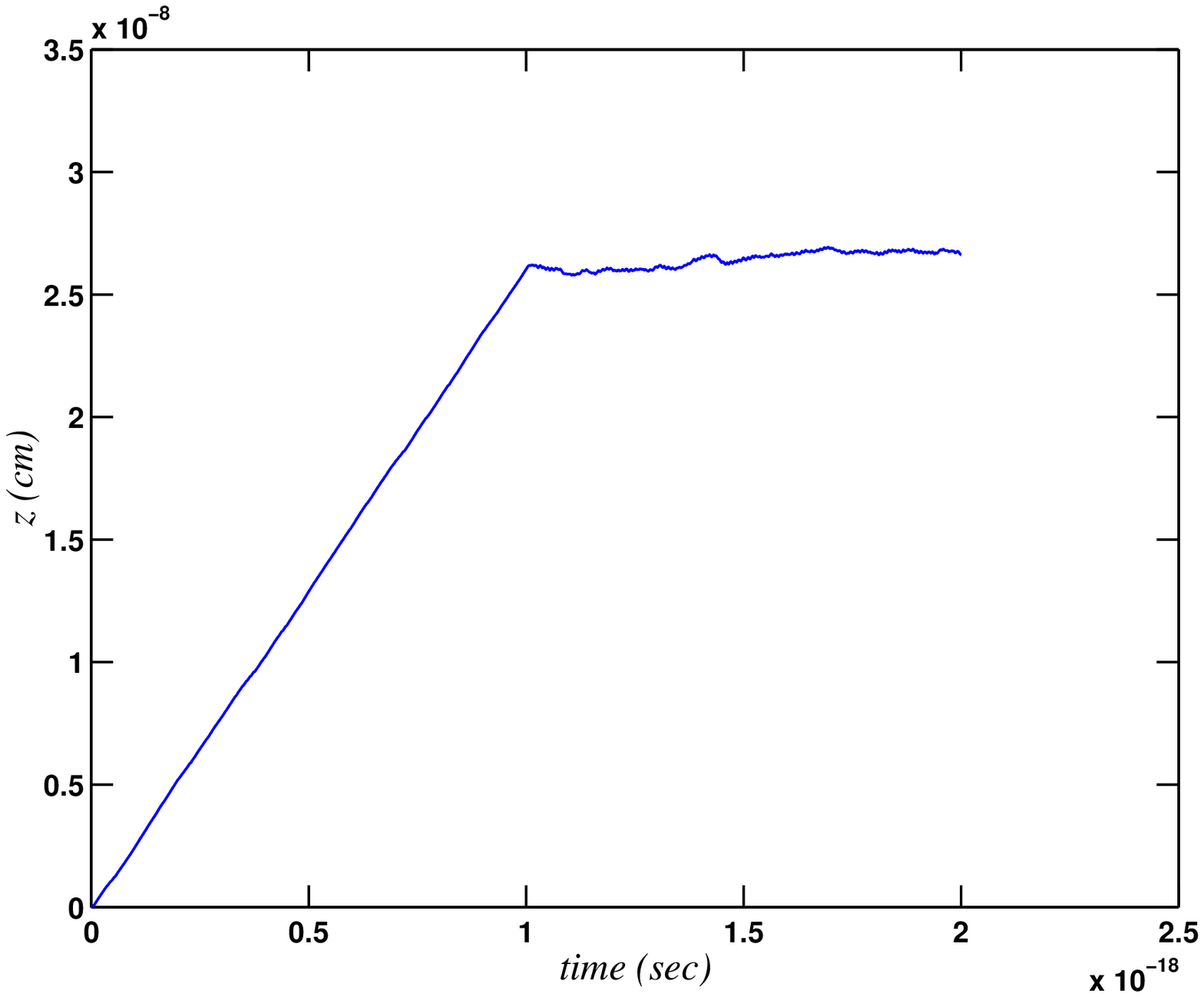} }
\begin{center}
 \begin{minipage}[t]{2.0in}
 {Figure 3. Drift in z-direction of Figure 1
 exhibiting non-inertial behavior as charge stops drifting
 when driving field is switched off.}
 \end{minipage}
\end{center}
\end{minipage}\hfill
\end{figure}

Now equation (\ref{eqmo}) by itself does not exhibit inertia. As
is illustrated in Figure 1 and more clearly in Figure 3, when the
driving field is switched off, the charge stops drifting and
resumes a zero-mean random walk in the frame of the calculation.
This non-inertial behavior is at distinct odds with relativity.

\section{The Connective transformation}

The non-inertial behavior of the massless charge equation of
motion (\ref{eqmo}) is in distinct violation of relativity. The
ZPF vacuum fields have the energy density frequency spectrum [7]

\be \rho(\omega)d\omega=\frac {\hbar\omega^3} {2\pi^2c^3}d\omega
\;\;\;. \ee

\ni The cubic frequency dependence of the ZPF spectrum endows it
with Lorentz invariance; all inertial frames see an isotropic ZPF.
A Lorentz transformation will cause a Doppler shift of each
frequency component, but an equal amount of energy is shifted into
and out of each frequency bin. When there are no fields above the
vacuum in an inertial frame, an observer in that frame should
expect to see a zero-mean random walk due to the isotropic ZPF.
Thus, in the example cited above, an observer in a frame co-moving
with the average motion of the charge just before the driving
field is switched off should expect to see continued zero-mean
zitterbewegung in his frame, whereas (\ref{eqmo}) produces
zero-mean motion in whatever frame the calculation is performed
in. To have a consistent theory, Lorentz covariance must be
restored.

We assume that (\ref{eqmo}) holds in the spacetime of the
particle. We further assume that in this spacetime, (\ref{eqmo})
is the equation of a null geodesic.  The curvature is defined by
the electromagnetic fields in the particle's history. Since the
charge is assumed to be massless and moving at the speed of light,
we cannot use proper time as the affine parameter of the geodesic
(proper time intervals vanish for null geodesics). However, normal
time serves well as an affine parameter. Thus we replace
(\ref{eqmo}) with

\be \frac {dp^\mu} {dt}+\frac {1} {m_*}\Gamma^\mu_{\nu \rho}p^\nu
p^\rho = 0 \;\;\;, \label{geo}\ee

\ni where

\be p^\mu=m_* c n^\mu \;\;\;,\ee

\ni and

\be n^\mu=(n^0,{\bf n}) \;\;\;.\ee

\ni $\Gamma^\mu_{\nu \rho}$ are the Christoffel symbols of the
second kind. We equate the connection terms (the terms containing
the Christoffel symbols) with the Lorentz force terms of equation
(\ref{lorentz}). That is, we set

\be \Gamma^\mu_{\nu \rho}p^\nu p^\rho = -\frac {q} {c} F^\mu_\nu
p^\nu \;\;\;. \label{connect} \ee

\ni These equations can be solved for the metric of the particle's
spacetime, though not uniquely.  The equations (\ref{connect})
actually define a class of metrics.  Further constraints are
required to select a particular solution from this class.  In
particular, the geodesic in the particle spacetime should be a
null curve as expected for a massless object.  Using
(\ref{connect}), the geodesic equation (\ref{geo}) can be written
as,

\be \frac {dn^j} {dt}=\frac {q} {m_*c} \left[F^j_\nu n^\nu -
F^0_\nu n^\nu \frac {n^j} {n^0} \right]+\frac {n^j} {n^0} \frac
{dn^0} {dt} \ee

\be \frac {dm_*} {dt}=\frac {q} {c} F^0_\nu \frac {n^\nu}
{n^0}-\frac {m_*} {n^0} \frac {dn^0} {dt} \label{mstareq}
\;\;\;.\ee

\noindent In general $n^0$ does not retain a value of unity, but
should change in a way that preserves the null curve property,

\be g_{\mu \nu} p^\mu p^\nu = 0 \label{null} \;\;\;. \ee

Note that (\ref{mstareq}), which is the zeroth equation of
(\ref{geo}), is an equation for the parameter $m_*$.  Thus $m_*$
is not a constant, but rather varies in response to applied
forces. The effect is to introduce time dilation (or Doppler
shifting) in the energy-momentum four vector analogous to the
gravitational redshift of General Relativity.

Also note that inertia is not assumed here by our requirement that
the particle travel on a null geodesic.  We do not restrict the
motion of the particle, rather we simply use its motion to define
the spacetime and metric that the particle sees.  It is only after
the transformation to Minkowski spacetime that inertial behavior
appears.

To find solutions of (\ref{connect}), we first consider an
infinitesimal region about the charge, and require

\be g_{\mu\nu}=\eta_{\mu\nu}+g_{\mu \nu , \rho} dx^\rho
\label{infg}\;\;\;,\ee

\ni where $\eta_{\mu\nu}$ is the flat spacetime metric,

\be \eta_{\mu\nu}=\left[\begin{array}{cccc}
  -1 & 0 & 0 & 0 \\
  0 & 1 & 0 & 0 \\
  0 & 0 & 1 & 0 \\
  0 & 0 & 0 & 1
\end{array}\right] \;\;\;.\ee

\ni The Christoffel symbols are then calculated in terms of the
derivatives $g_{\mu \nu , \rho}$ and substituted into
(\ref{connect}), resulting in an underdetermined set of equations
for the metric derivatives.  One solution for the metric tensor
(valid for an infinitesimal patch about the position of the
charge) is,

\begin{eqnarray}
g_{00} & = & -1+\frac {q} {m_*c^2n_0}(E_x dx+E_y dy+E_z dz)=-1+F_L dt \nonumber \\
g_{01} & = & -\frac {q} {2m_*c^2n_0}(E_x cdt+B_z dy-B_y dz)=-\frac {1} {2} F_x dt \nonumber \\
g_{02} & = & -\frac {q} {2m_*c^2n_0}(E_y cdt+B_x dz-B_z dx)=-\frac {1} {2} F_y dt \nonumber \\
g_{03} & = & -\frac {q} {2m_*c^2n_0}(E_z cdt+B_y dx-B_x dy)=-\frac {1} {2} F_z dt \nonumber \\
g_{11} & = & g_{22} = g_{33} = 1  \nonumber \\
g_{12} & = & g_{13} = g_{23} = 0  \label{infmetric}
\;\;\;,\end{eqnarray}

\ni with the remaining terms given by the symmetry of
$g_{\mu\nu}$. Here we use the normalized force $F=\mbox{{\em
force}}/m_*cn_0$, and $F_L$ is the longitudinal component with
respect to the direction of particle motion.  The second
equalities in (\ref{infmetric}) hold only on the particle
geodesic.

The metric $g_{\mu\nu}$ implies a transformation to Minkowski
spacetime. The transformation $C_\mu^\nu$ from the particle's
spacetime to Minkowski spacetime is related to the metric
$g_{\mu\nu}$ by $g_{\mu\nu}=C^\rho_\mu C^\sigma_\nu
\eta_{\rho\sigma}$, or

\be g=C\cdot\eta\cdot\tilde{C} \label{Ctog} \;\;\;,\ee

\ni where $\tilde{C}$ is the transpose of $C$.  For the metric
(\ref{infmetric}) we find (for transformations along the particle
geodesic),

\be C_{\mbox{{\em infinitesimal}}}=\left[\begin{array}{cccc}
  1-F_L dt & \frac {1} {4} F_x dt & \frac {1} {4} F_y dt & \frac {1} {4} F_z dt \\
  -\frac {1} {4} F_x dt & 1 & 0 & 0 \\
  -\frac {1} {4} F_y dt & 0 & 1 & 0 \\
  -\frac {1} {4} F_z dt & 0 & 0 & 1
\end{array}\right] \;\;\;.\ee

A general finite transformation can be obtained by repeated
application of the infinitesimal one.  In the infinite limit this
yields,

\begin{eqnarray}
C_0^0(t) & = & \frac {1} {F_T} \exp \left( -\frac {1} {4} \int_0^t
F_L(t^\prime) dt^\prime \right) \left[ F_T \cos \left(\frac {1}
{4} \int_0^t F_T(t^\prime) dt^\prime\right)-F_L \sin \left(\frac
{1} {4} \int_0^t F_T(t^\prime) dt^\prime\right)\right] \nonumber \\
C_0^i(t) & = & -\frac {1} {F_T} \exp \left( -\frac {1} {4}
\int_0^t F_L(t^\prime) dt^\prime \right) F_i \sin \left(\frac
{1} {4} \int_0^t F_T(t^\prime) dt^\prime\right) \nonumber \\
C_i^i(t) & = & \frac {1} {F^2 F_T} \left\{F_T (F^2-F_i^2) \right. \nonumber \\
& + & \left. \exp \left( -\frac {1} {4} \int_0^t F_L(t^\prime)
dt^\prime \right) F_i^2\left[ F_T \cos \left(\frac {1} {4}
\int_0^t F_T(t^\prime) dt^\prime\right)+F_L \sin \left(\frac
{1} {4} \int_0^t F_T(t^\prime) dt^\prime\right)\right]\right\} \nonumber \\
C_i^j(t) & = & \frac {F_i F_j} {F^2 F_T} \left\{-F_T \right. \nonumber \\
& + & \left. \exp \left( -\frac {1} {4} \int_0^t F_L(t^\prime)
dt^\prime \right) \left[ F_T \cos \left(\frac {1} {4} \int_0^t
F_T(t^\prime) dt^\prime\right)+F_L \sin \left(\frac {1} {4}
\int_0^t F_T(t^\prime) dt^\prime\right)\right]\right\}
\label{Ctrans} \end{eqnarray}

\ni where $i,j=1,2,3$, $C_i^0=-C_0^i$, $C_j^i=C_i^j$, and $F_T$ is
the magnitude of the transverse component of the normalized force
(transverse with respect to the direction of particle motion). The
metric $g_{\mu\nu}$ of the particle's spacetime (on the geodesic)
is obtained through (\ref{Ctog}),

\begin{eqnarray}
g_{00}(t) & = & -\frac {1} {F_T} \exp \left( -\frac {1} {2}
\int_0^t F_L(t^\prime) dt^\prime \right) \left[ F_T \cos
\left(\frac {1} {2} \int_0^t F_T(t^\prime) dt^\prime\right)-F_L
\sin \left(\frac
{1} {2} \int_0^t F_T(t^\prime) dt^\prime\right)\right] \nonumber \\
g_{0i}(t) & = & -\frac {1} {F_T} \exp \left( -\frac {1} {2}
\int_0^t F_L(t^\prime) dt^\prime \right) F_i \sin \left(\frac
{1} {2} \int_0^t F_T(t^\prime) dt^\prime\right) \nonumber \\
g_{ii}(t) & = & \frac {1} {F^2 F_T} \left\{F_T (F^2-F_i^2) \right. \nonumber \\
& + & \left. \exp \left( -\frac {1} {2} \int_0^t F_L(t^\prime)
dt^\prime \right) F_i^2\left[ F_T \cos \left(\frac {1} {2}
\int_0^t F_T(t^\prime) dt^\prime\right)+F_L \sin \left(\frac
{1} {2} \int_0^t F_T(t^\prime) dt^\prime\right)\right]\right\} \nonumber \\
g_{ij}(t) & = & \frac {F_i F_j} {F^2 F_T} \left\{-F_T \right. \nonumber \\
& + & \left. \exp \left( -\frac {1} {2} \int_0^t F_L(t^\prime)
dt^\prime \right) \left[ F_T \cos \left(\frac {1} {2} \int_0^t
F_T(t^\prime) dt^\prime\right)+F_L \sin \left(\frac {1} {2}
\int_0^t F_T(t^\prime) dt^\prime\right)\right]\right\}
\label{metric} \end{eqnarray}

\ni with the remaining terms given by symmetry.  As we said, the
solution (\ref{Ctrans}) and (\ref{metric}) is not unique, but it
is in some sense the simplest one as it yields the equality of the
connection terms and the Lorentz forces in a term-by-term manner.
We will show in the next section that this metric has several
desirable properties, particularly the generation of both inertia
and spin.  It also implies a many-particle theory with pair
creation and annihilation.

One interesting aspect of (\ref{metric}) is that it depends on the
particle history through the time integrations.  This means that
the spacetime of a particle is unique to it; two particles at the
same place at the same time will be in different spacetimes if the
history integrals of (\ref{metric}) differ (as they generally
will).

A second interesting point is that the integration over
longitudinal forces appears within exponentials whereas the
integration over transverse forces appears within oscillatory
trigonometric functions.  The latter may account for pair creation
and annihilation.  The former produces time dilation that, as we
shall soon see, yields the effect we know as inertia.

\section{Inertia}

We have shown in Figure 3 that the massless charge equation of
motion (\ref{eqmo}) does not contain the effect of inertia.  The
formulation of the preceding section restores Lorentz covariance
to the theory through the geodesic equation (\ref{geo}), and thus
we expect that the well-known effects of inertia should appear.
Two of the simplest manifestations of inertia are hyperbolic
motion of a charge in the presence of a uniform electric field and
uniform motion following an impulse.  We now demonstrate these
effects.

Figure 4 displays the component of the motion of a charge in the
direction of a uniform electric field applied above the ZPF,
obtained by solving (\ref{geo}) using the metric (\ref{metric})
and applying the Connective transformation to view the result in
Minkowski spacetime.  The solid curve is the result of the
Connectivity simulation.  The circles lie on the hyperbola defined
by a massive charge undergoing uniform acceleration in special
relativity.  The agreement is striking.  Viewed in Minkowski
spacetime, the massless charge is seen to accelerate
hyperbolically as though it has inertia.  Figure 5 is a similar
depiction for the case in which an impulse has been applied to the
charge, and the charge is observed in Minkowski spacetime to
continue in uniform motion following the impulse.  Here the
circles lie on a straight line, indicating that the particle
travels at a constant speed following the impulse.  Note that it
is the average motion of the charge that moves uniformly.
Deviations about the average motion are apparent and are, in fact,
zitterbewegung driven by the ZPF.  The average motion defines a
timelike curve, as expected of a massive particle.  (That the
metric (\ref{metric}) produces instantaneous motion on a {\em
null} curve needs to be verified since our equations {\em assume}
a massless particle.  From our numerical simulations, this appears
to be at least approximately true for the single-particle regime.)

\begin{figure}[h]
\begin{minipage}[t]{3.in}
\epsfysize=3.in
\centerline{\epsffile{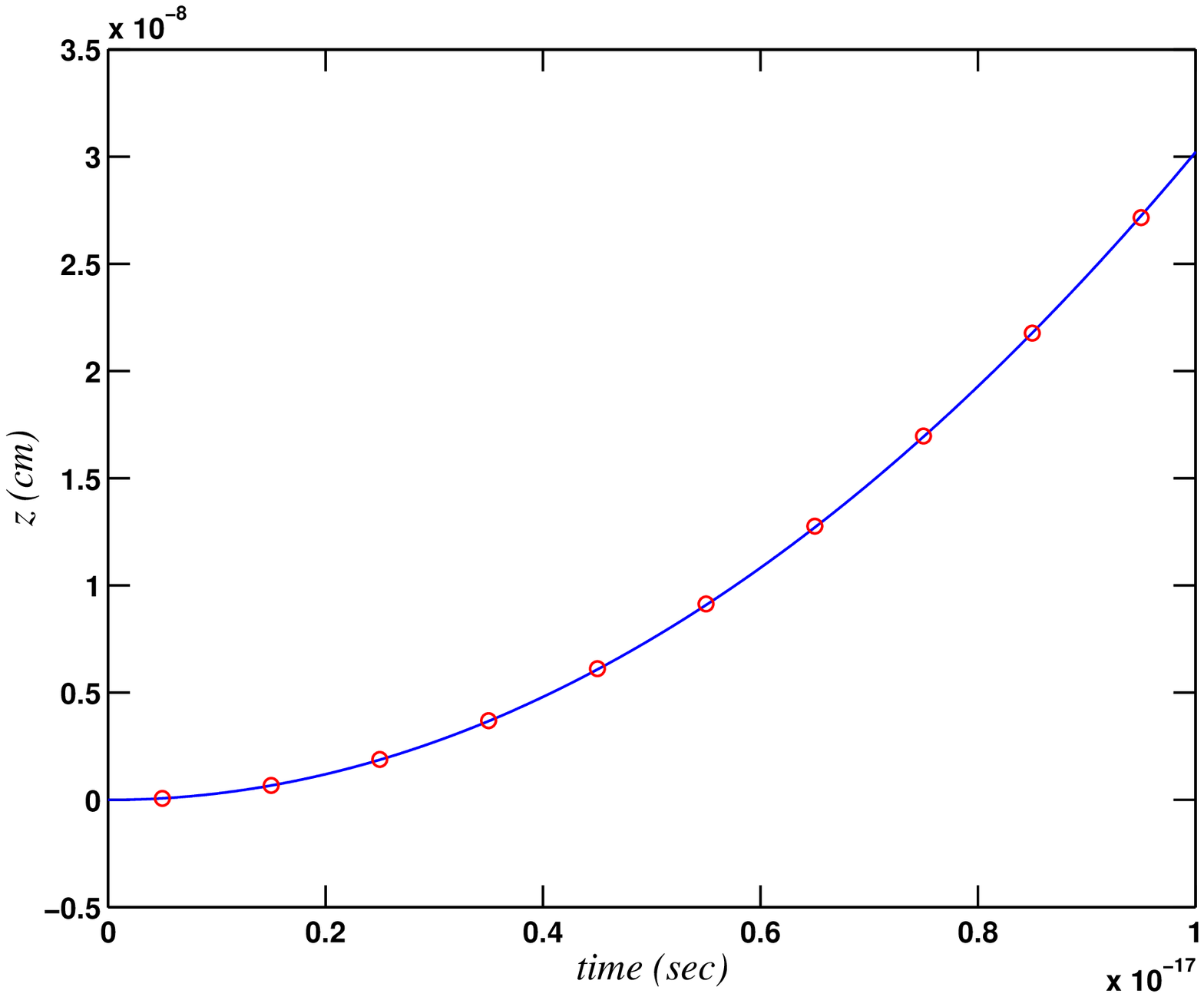} }
\begin{center}
 \begin{minipage}[t]{3.in}
{Figure 4.  Massless charge motion in a uniform electric field
(plus ZPF), obtained using Connectivity (solid curve), compared to
the hyperbolic motion of a massive charge in special relativity
(circles).}
 \end{minipage}
\end{center}
\end{minipage}
\hfill
\begin{minipage}[t]{3.in}
\epsfysize=3.in
\centerline{\epsffile{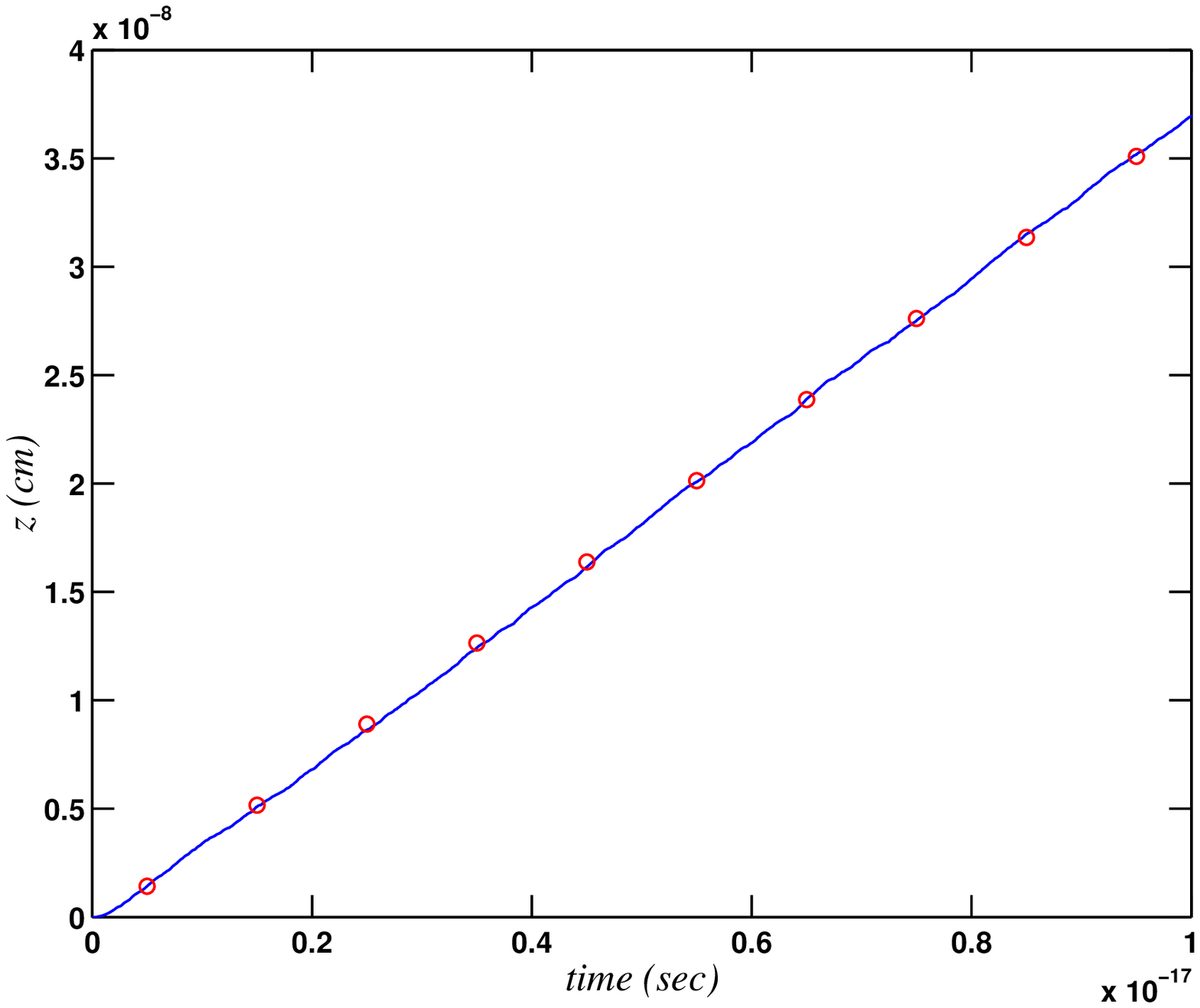} }
\begin{center}
 \begin{minipage}[t]{3.in}
 {Figure 5. Massless charge motion in the ZPF following an impulse, obtained
 using Connectivity (solid curve), compared to constant
 speed motion (circles).}
 \end{minipage}
\end{center}
\end{minipage}
\hfill
\end{figure}

In the simulations of Figures 4 and 5, we were careful to require
that the charge was never energized enough for particle pair
creation to occur.  This was done by initiating the simulation
with a sufficiently large value of $m_*$.  As is apparent in
(\ref{Ctrans}), if the particle is energized sufficiently by the
ZPF plus driving field, the Connective transformation can cause
positive time intervals in the particle's spacetime to be
transformed to negative time intervals in Minkowski spacetime.
This negative-time transport of the particle can be viewed as the
positive-time transport of its antiparticle.  The multi-particle
regime was avoided here so that the inertia-generating aspect of
Connectivity could be demonstrated without complication.

\section{Discussion}

The premise of the Connectivity approach is that charges are
massless and move on null geodesics in a spacetime whose curvature
is defined by equating the forces on the charge with the
connection terms in the geodesic equation.  These forces include
those of the underlying vacuum fields.  The metric of this
spacetime defines a transformation to Minkowski spacetime, wherein
the charge is observed to exhibit inertial behavior.

The simple fact that the Connectivity premise yields the effects
of inertia from massless charges is strong motivation to seriously
consider other implications of the theory, especially with regard
to its potential to shed light on the basis of the quantum theory.

When the forces acting on the charge are the Lorentz forces due to
the electromagnetic vacuum fields (the ZPF), these drive the
charge in zitterbewegung motion at the speed of light, in
agreement with the speed-of-light eigenvalues of the Dirac theory.
When the charge moves with a large average velocity in some
direction, the zitterbewegung motion extends to a quasi-helical
motion that may be the basis of particle spin.  This spin is
undetermined until ``measured'' by applying a field that aligns
the zitterbewegung into helical motion, which will either be
oriented with positive or negative helicity (spin up or spin
down). Further work will be required to determine whether it is
possible for this ``spin'' to manifest the spin correlation
properties of the quantum theory.

It is implied by the Connective transformation (\ref{Ctrans}) that
the charge can be observed in Minkowski spacetime to be travelling
backward in time, or rather that its antiparticle is travelling
forward in time.  Thus we have the basis for a many particle
theory that manifests the degrees of freedom of the Dirac theory
(positive and negative energy states, each with spin up/down
configurations).

There are a number of directions for further research that are
implied by the Connectivity premise.  These include studying the
correspondence of Connectivity with quantum theory, the theory of
electromagnetic radiation, and with gravitation theory.

For example, we suspect that it might be possible to derive the
Schr\"{o}dinger and Dirac equations as stochastic equations
obtained from the ensemble average of the Connective
transformation applied to the geodesic equation (\ref{geo}) in the
single- and multi-particle regimes respectively.

Since the interpretation of Connectivity is that electromagnetic
fields define a curvature of spacetime, it may be possible to show
that radiation is related to the distortion of spacetime required
to connect the views of different frames. This might then have
implications for the problem of the stability of atoms.

Exploring the relationship between the particle spacetime of
Connectivity (which generates inertia) to the spacetime of General
Relativity may shed light on Einstein's postulate of the
Equivalence Principle.  Among the things assumed in General
Relativity are {\em i.}) the Equivalence Principle, {\em ii.}) a
spacetime with a metric which is well defined even in the absence
of matter and energy, {\em iii.}) matter and energy, which are
already assumed to have well defined rest masses and only distort
the metric.  Given these (with appropriate topology and boundary
conditions), the inertial frames are well defined. However, the
standard version of relativity does not explain why the
Equivalence Principle holds in the first place, how the metric can
be well defined in an empty spacetime (how do we measure motion
without reference points?), and what are the rest masses of the
individual particles (or the weight of an object held up against
gravity). In each case, the solution of these problems may lie in
considering the effect of the vacuum as well as that of the matter
and energy above the vacuum.  Even in an ``empty'' spacetime,
vacuum fields can provide spacetime events to which motion can be
referred and acceleration with respect to the vacuum can be
detected. Secondly, according to Connectivity, interaction with
the vacuum endows individual particles with inertia and a
corresponding effective mass, possibly providing a means of
determining the individual rest masses. Finally, if we can derive
inertial effects from interaction with the vacuum, it seems
plausible that we should be able to similarly derive gravitational
effects, thereby yielding the equivalence principle and the
Minkowski metric. Explicit reference to (and interaction with) a
real vacuum may provide reasonable explanations for the
fundamental assumptions of gravitation theory as represented in
General Relativity.

\section{Appendix A: ZPF Realizations}

A convenient way to explore the properties of the massless charge
equation of motion (\ref{eqmo}) and the Connective transformation
(\ref{Ctrans}) is through numerical simulation.  This requires the
generation of ZPF realizations.  We use the prescription of Ibison
and Haisch [8],

\be {\bf E}({\bf r},t)=\mbox{{\em Re}}\int d^3k \sum_{\lambda=1}^2
\hat\epsilon_{k,\lambda}
\frac{\sqrt{\hbar\omega}}{2\pi}(u_{k,\lambda}+iv_{k,\lambda})\exp\left[i({\bf
k}\cdot {\bf r}-\omega t)\right] \label{realize1} \ee

\be {\bf B}({\bf r},t)=\mbox{{\em Re}}\int d^3k \sum_{\lambda=1}^2
\hat k \times \hat\epsilon_{k,\lambda}
\frac{\sqrt{\hbar\omega}}{2\pi}(u_{k,\lambda}+iv_{k,\lambda})\exp\left[i({\bf
k}\cdot {\bf r}-\omega t)\right] \;\;\;. \label{realize2} \ee

\ni Here $u_{k,\lambda}$ and $v_{k,\lambda}$ are Gaussian random
numbers with zero mean and unit variance, ${\bf k}$ is the wave
vector of a plane wave component of the ZPF, and $\lambda$ is the
polarization index.  Gaussian electromagnetic units have been
used.

In generating random realizations of the ZPF, two practical
considerations are forced upon us.  Although the ZPF may
theoretically extend to infinite frequencies, we must assume a
cutoff frequency to build a numerical realization.  There are
physical reasons for applying a cutoff frequency, not in the ZPF
{\em per se}, but in the charge's response to the ZPF.  This is
related to our spin interpretation of the orbital angular momentum
of the helical ZPF-driven motion of the charge.  The value of this
angular momentum is dependent on the presumed cutoff frequency,
and for the electron this implies a cutoff in the vicinity of the
electron Compton frequency.  The simplest way to visualize this
cutoff is to assume that the charge has a finite spatial extent
and is not, in fact, a point singularity.  It is then only
influenced by the portion of the ZPF spectrum with wavelengths
longer than the charge distribution size.  The second
consideration affecting our random realizations is related to the
size of the realization. Ideally one would generate a complete
space-time realization using (\ref{realize1}) and
(\ref{realize2}), but since we want to consider frequencies up to
a cutoff frequency near the electron Compton frequency and also
consider comparatively long time durations, computational memory
storage becomes an issue. In order to be able to run the
simulations on a very modest computer, we make the following
compromise.  We generate a time-history realization at the point
${\bf r}=0$, but apply this realization to the charge no matter
where it is.  Since the realization has the correct spectral
properties in frequency, this approximation can be expected to
correctly give most of the behaviors we seek to demonstrate, but
we must realize that certain spatial correlation aspects may not
be correctly represented. Thus, for example, we see an orbital
angular momentum emerging in our realizations of the charge motion
that we interpret to be spin, but if one wants to simulate spin
correlation effects for comparison to the predictions of the
quantum theory, the spatial aspect of (\ref{realize1}) and
(\ref{realize2}) should be included.

\rule[0pt]{0pt}{12pt}
\begin{center}
{\bf Acknowledgement}
\end{center}

\rule[0pt]{0pt}{10pt}

The work of Dr.~Nickisch was supported in part by the California
Institute for Physics and Astrophysics.

\rule[0pt]{0pt}{12pt}
\begin{center}
{\bf References}
\end{center}

\rule[0pt]{0pt}{10pt}

\ni [1] B.~Haisch, A.~Rueda, and H.~E.~Puthoff, Phys.~Rev.~A {\bf
49}, 678 (1994).

\ni [2] A.~Rueda and B.~Haisch, Foundations of Physics {\bf 28},
1057 (1998).

\ni [3] E.~Schr\"{o}dinger, Sitzungsb.~Preuss.
Akad.~Wiss.~Phys.-Math.~K1.~{\bf 24}, 418 (1930); {\bf 3}, 1
(1931).

\ni [4] P.~A.~M.~Dirac, {\em The Principles of Quantum Mechanics},
(Clarendon, Oxford), 4th edition, p.~262 (1958).

\ni [5] K.~Huang, Am.~J.~Physics {\bf 20}, 479 (1952).

\ni [6] A.~O.~Barut and N.~Zanghi, Phys.~Rev.~Lett. {\bf 52}, 2009
(1984).

\ni [7] L.~de la Pe\~{n}a and M.~Cetto, {\em The Quantum Dice: An
Introduction to Stochastic Electrodynamics}, (Klewer Acad.~Publ.),
chap.~4 (1996).

\ni [8] M.~Ibison and B.~Haisch, Phys.~Rev.~A {\bf 54}, 2737
(1996).

\end{document}